\documentclass[11pt,twoside,a4paper,cmspaper,final,collab]{cms-tdr}
\def\svnVersion{f75c307-D}\def\svnDate{2026/08/02}\def\cmsCernNoTag{CERN-EP-2026-231}\def\cmsCernDate{\today}\def\cmsMessage{Submitted to Physical Review Letters}
\begin{document}\cmsNoteHeader{HIN-25-020}

\newcommand{\pp}{\ensuremath{\Pp\Pp}\xspace}
\newcommand{\pPb}{\ensuremath{\Pp\text{Pb}}\xspace}
\newcommand{\OO}{\ensuremath{\text{OO}}\xspace}
\newcommand{\raa}{\ensuremath{R_\text{AA}}\xspace}
\newcommand{\roo}{\ensuremath{R_\text{OO}}\xspace}
\newcommand{\dSigdPtAA}{\ensuremath{\frac{\rd \sigma^{\mathrm{AA}}}{\rd \pt}}\xspace}
\newcommand{\dSigdPtpp}{\ensuremath{\frac{\rd \sigma^{\pp}}{\rd \pt}}\xspace}
\newcommand{\dNdEta}{\ensuremath{\rd N_{\mathrm{ch}}/\rd \eta}\xspace}

\newlength\cmsFigWidth
\ifthenelse{\boolean{cms@external}}
{   
\setlength{\cmsFigWidth}{0.99\linewidth}
}
{
\setlength{\cmsFigWidth}{0.7\textwidth}
}

\cmsNoteHeader{HIN-25-020}

\title{Observation of enhanced baryon production using strange hadrons in oxygen-oxygen collisions}

\author[cern]{CMS Collaboration}

\abstract{
A hot deconfined medium known as the quark-gluon plasma (QGP) is created in collisions of ultrarelativistic heavy nuclei.  In these systems, hadron production is strongly modified relative to proton-proton (\pp) interactions.
In the hadron transverse momentum region $2<\pt<5\GeV$, collective radial flow and quark coalescence are important mechanisms that can produce such modifications.  While these effects are well-established for collisions of large ions, such as lead, it is unknown if they are present in systems produced using smaller ions.  To address this question, the \pt spectra of two strange hadrons, \PKzS and \PGL, are measured in oxygen-oxygen (\OO) and \pp collisions at a center-of-mass energy per nucleon pair of 5.36\TeV.  The measurements are performed using the CMS detector at the LHC, and the \OO and \pp datasets correspond to integrated luminosities of 6.8\nbinv and 1.1\pbinv, respectively.  The spectral modification of \PKzS mesons in \OO collisions relative to \pp collisions is found to be similar to that of inclusive charged hadrons. Conversely, the \PGL-to-\PKzS cross section ratio exhibits a pronounced enhancement in \OO collisions when compared to \pp collisions. The results provide the first observation of a baryon-to-meson enhancement in \OO collisions and suggest the formation of a deconfined medium, offering insight into the roles of strangeness, radial flow, and quark coalescence in QGP.}

\hypersetup{%
pdfauthor={CMS Collaboration},%
pdftitle={Observation of enhanced baryon production using strange hadrons in oxygen-oxygen collisions},%
pdfsubject={CMS},%
pdfkeywords={CMS, heavy ions, quark-gluon plasma, particle suppression, nuclear modification factor, strange hadrons, strangeness enhancement, baryon enhancement, baryon-to-meson ratio}}

\maketitle 

A hot medium of deconfined quarks and gluons, known as the quark-gluon plasma (QGP), is created in collisions of ultrarelativistic heavy nuclei~\cite{Busza:2018rrf,Shuryak:1980tp}.  As this medium cools down, it transitions to a hadron gas phase through a process known as hadronization.  Although hadronization is not well understood from first-principles quantum chromodynamics (QCD) calculations, it can be described to various degrees of accuracy with phenomenological models~\cite{Andersson:1983ia,Webber:1983if,Aichelin:2008mi}.  The experimental study of the resulting hadrons gives insight into in-medium hadronization mechanisms and QGP dynamics.

A unique mechanism present within QGP, which has a temperature well above the masses of light quarks, is the thermal production of strange quarks via gluon splitting~\cite{Koch:1986ud}.  These strange quarks hadronize with other partons to produce additional strange hadrons, such as \PKzS mesons and \PGL baryons.  This strangeness enhancement effect was one of the first experimental signatures hinting at QGP production at the CERN SPS~\cite{WA97:1999uwz,NA49:2002pzu,NA57:2010tnk} and the BNL RHIC~\cite{STAR:2007cqw}.  Another example of QGP-induced modification to hadron production is the so-called baryon anomaly.  Deconfined quarks and gluons can coalesce into baryons at a higher rate than collision systems without QGP formation~\cite{Fries:2008hs,Minissale:2015zza}.  When combined with the effects of radial flow~\cite{Hung:1997du}, where pressure gradients within the QGP impart additional transverse momentum (\pt) to the deconfined partons, the resulting baryons also tend to have slightly higher \pt.  Experimentally, the baryon anomaly manifests as an enhancement of baryon yields relative to meson yields in the region of $2<\pt<5\GeV$. The strangeness enhancement and baryon anomaly effects have been well-studied in heavy ion collisions at RHIC~\cite{PHENIX:2003tvk,STAR:2006egk,STAR:2007zea,STAR:2011fbd} and the CERN LHC~\cite{ALICE:2013cdo,ALICE:2013xmt}.

These effects have also been observed in proton-proton (\pp)~\cite{ALICE:2016fzo,ALICE:2019avo,ALICE:2025xji}, proton-lead (\pPb)~\cite{ALICE:2013wgn,ALICE:2015mpp}, deuteron-gold~\cite{Zhu:2009uq}, and even photon-ion collisions~\cite{ATLAS:2025tof}.  In general, these studies find that the production of strange hadrons and baryons increases with event activity~\cite{ALICE:2016fzo}.  One interpretation of these data is that a QGP is also produced in these systems; a hypothesis that is also supported by examination of other observables such as hydrodynamic flow~\cite{Nagle:2018nvi,Weller:2017tsr}.  On the other hand, these observations have prompted discussion regarding the necessity of QGP production to explain hadronization modifications, with other explanations examining the connection of multiparton interactions with strangeness and baryon production~\cite{OrtizVelasquez:2013ofg,Bierlich:2015rha,Bierlich:2014xba}.  In all of these systems, hadronization modifications are most prominent when selecting collisions with high event activity, which has then raised questions regarding possible selection biases on the observables of interest.

In 2025, the LHC collided oxygen ions (having mass number $\text{A}=16$) at a center-of-mass energy per nucleon pair of $\sqrtsNN = 5.36\TeV$.  Examination of hadron production in oxygen-oxygen (\OO) collisions has already found signatures of QGP production~\cite{CMS:2025bta,CMS:2025tga,CMS:2025tga,ALICE:2025luc,ATLAS:2025nnt,ALICE:2026zck,CMS:2026qbb,LHCb:2026gwz}, but potential modifications to strange hadron and baryon yields have yet to be quantified.  Since \OO collisions produce charged-particle multiplicities comparable to high-multiplicity \pPb collisions~\cite{CMS:2026sai}, this dataset provides an opportunity to search for strangeness enhancement and baryon anomaly effects in intermediately sized collision systems without the complications that arise from event activity selections. 

This Letter presents the first measurement of identified strange hadrons in \OO collisions at $\sqrtsNN = 5.36\TeV$ using data collected by the CMS experiment. A baseline measurement using \pp data collected in 2024 at the same collision energy is also presented. First, the \pt-differential invariant cross sections of \PKzS and \PGL particles are presented.  Second, the nuclear modification factor is calculated using the formula 
\begin{equation}
\raa = \frac{1}{\text{A}^2}\left[\left(\dSigdPtAA\right)\bigg{/} \left(\dSigdPtpp\right)\right],
\label{eq1}
\end{equation}
where $\sigma^{\mathrm{AA}}$ ($\sigma^{\pp}$) is the cross section for identified-particle production in \OO (\pp) collisions.  Finally, the \PGL-to-\PKzS cross section ratio measured in \OO and \pp collisions is used to probe \pt-dependent baryon enhancement. Tabulated results are provided in the HEPData record for this analysis~\cite{hepdata}.

The CMS apparatus~\cite{CMS:2008xjf,CMS:2023gfb} is a multipurpose, nearly hermetic detector, designed to trigger on~\cite{CMS:2020cmk,CMS:2016ngn,CMS:2024aqx} and identify electrons, muons, photons, and hadrons~\cite{CMS:2020uim,CMS:2018rym,CMS:2014pgm}. A global ``particle-flow'' algorithm~\cite{CMS:2017yfk} aims to reconstruct all individual particles in an event, combining information provided by the all-silicon inner tracker and by the crystal electromagnetic and brass scintillator hadron calorimeters, operating inside a 3.8\unit{T} superconducting solenoid, with data from the gas-ionization muon detectors embedded in the flux-return yoke outside the solenoid. The silicon tracker~\cite{Phase1Pixel} measures charged particles up to $\abs{\eta} = 3.0$ with typical resolutions of 2.0\% in \pt and 10--75\mum in the transverse impact parameter~\cite{DP-2020-049} for non-isolated particles of $1 < \pt < 100\GeV$.  The hadron forward (HF) calorimeter uses steel as an absorber and quartz fibers as the sensitive material. The two halves of the HF calorimeter are located 11.2\unit{m} from the interaction region, one on each end, and together they provide coverage in the range $3.0 < \abs{\eta} < 5.2$.

Monte Carlo (MC) samples are utilized to evaluate detector performance and calculate correction factors.  The \PYTHIA 8.311 event generator~\cite{Bierlich:2022pfr} with the CP5 tune~\cite{CMS:2019csb} is used to model \pp collisions. To study \OO collisions, hard-scattering events generated with \PYTHIA are overlaid with \OO~\HIJING v1.383~\cite{Wang:1991hta} events to simulate underlying particle production. Detector response is simulated with the \GEANTfour~\cite{Agostinelli:2002hh} program.  After event generation, an event-by-event weight matches the distribution of the $z$ position of reconstructed primary vertices (PV)~\cite{CMS:2014pgm} in simulation to those in data.

A trigger that required only two coincident colliding bunches in the detector was used to collect the \pp data. The \OO data were recorded with a trigger searching for at least one energy deposit of at least 7\GeV in either HF calorimeter. The total integrated luminosities are $1.06\pm0.03\pbinv$ and $6.8\pm0.3\nbinv$ for the \pp and \OO samples, respectively. In each collision system, these values are determined from two independent measurements: a preliminary analysis of van der Meer scan~\cite{vanderMeer:296752,CMS:2025rzq} data and a second luminosity determination that compares observed $\PZ$ boson yields to a well-calibrated reference dataset~\cite{CMS:2023pad}.

The average pileup, defined as the mean number of interactions per bunch crossing ($\mu$), was 4.7 and 0.14 for the \pp and \OO datasets, respectively. A pileup correction was derived by studying the \PKzS and \PGL yield per event as a function of the number of reconstructed PV for different pileup conditions. The measured spectra are corrected to the reference corresponding to the lowest $\mu$ conditions. This results in an increase of the \PKzS and \PGL yields in \pp collisions by 4--5\% for $\pt< 1\GeV$ and an increase of 1--2\% above 3\GeV, depending on the particle of interest. Conversely, the \OO yields are reduced by 3--4\% for $\pt<1\GeV$  and 1\% for $\pt>3\GeV$.

Collision events must have at least one reconstructed PV within 15\unit{cm} of the detector center in the $z$ direction. The \OO events are required to produce energy deposits above 4\GeV in two or more detector elements on each side of the HF calorimeter to suppress contributions from beam background and ultraperipheral collisions~\cite{Baltz:2007kq}.  The effect of this selection was evaluated by varying the requirement to request at least one energy deposit over 13\GeV in each HF side; the resulting difference in the measured spectra is below 0.5\%.  Beam contamination from $^{16}$O dissociation processes~\cite{Pshenichnov:1997un} was found to be negligible by studying the time dependence of event activity distributions~\cite{Das:2025rqb}.

The \PKzS and \PGL candidates are reconstructed from pairs of oppositely charged tracks with track $\pt > 0.1\GeV$ and $\abs{\eta} < 2.4$, which are fitted to a common secondary vertex (SV) using a Kalman filter~\cite{Fruhwirth:1987fm}. Charged-particle tracks are reconstructed with the standard \pp tracking algorithms using hits in the silicon tracker~\cite{CMS:2014pgm}. Selected tracks are required to contain at least three valid hits, and the quality of the track fit must have a $\chi^{2}$ per degree of freedom less than 7. The transverse and longitudinal impact parameter significances, defined as the impact parameters normalized by their uncertainties, are calculated with respect to the PV. Both significances must exceed unity to enhance the selection of tracks originating from displaced SVs. In addition, the distance of closest approach significance of the two daughter tracks to each other must be less than one standard deviation in both the transverse plane and longitudinal direction.

The \PKzS and \PGL candidates are reconstructed through the $\PGpp\PGpm$ and $\Pp \PGpm$ decay channels, respectively. The charge-conjugate $\PAGL \to \PAp \PGpp$ process is not analyzed. The candidates are required to have rapidity $\abs{y}<1.0$, and $\pt > 0.5~(0.7)\GeV$ for \PKzS (\PGL), respectively. The invariant mass is computed by assigning the corresponding particle mass hypotheses to the daughter tracks, and candidates are selected within the mass windows $0.433 < m_{\PGpp\PGpm} < 0.563\GeV$ for \PKzS and $1.092 < m_{\Pp \PGpm} < 1.140\GeV$ for \PGL. To ensure good signal purity, candidates are required to satisfy the following selection criteria. The $\chi^{2}$ per degree of freedom of the SV fit must be smaller than 7. For each candidate, the associated PV is chosen by minimizing the three-dimensional (3D) distance of closest approach between the candidate momentum vector and the PV position. The 3D decay length and pointing angle, defined as the angle between the reconstructed candidate momentum vector and the line connecting the PV and SV, are then computed. The pointing angle is required to satisfy $\cos\theta > 0.999$, and the decay length significance must exceed 5 standard deviations.

Signal yields are extracted by fitting the invariant mass spectra of the \PKzS and \PGL candidates in intervals of candidate \pt. The signal component is modeled using a double-Gaussian function, while the combinatorial background and residual contributions from misreconstructed tracks are described by a third-order polynomial.  The signal yields are corrected for the detector acceptance, reconstruction efficiency, and selection efficiency using simulated events. The reported \PKzS and \PGL yields correspond to primary particles, defined as those produced directly in the collision or from decays of particles with a mean proper lifetime less than 1\unit{cm}. The primary \PGL baryon definition thus includes contributions from \PGSz decays but excludes those originating from weak decays of heavier strange baryons (\PGX and \PGO).  

For the \PGL measurement, the feed-down contribution from weak decays of \PGX and \PGO baryons is estimated using simulated events, in which the generated \PGX and \PGO spectra are weighted to match the corresponding measurements from the ALICE Collaboration~\cite{ALICE:2025xji}, assuming the yields of \PGXz and \PGXm are equal. For \OO collisions, an additional multiplicative correction factor is applied to account for the expected modification of strange hadron production relative to \pp collisions, following a strategy similar to that adopted in Ref.~\cite{CMS:2025bta}. The correction is derived from the measured evolution of the $\PGX/\PGL$ and $\PGO/\PGL$ yield ratios~\cite{ALICE:2016fzo} as a function of charged-particle multiplicity density, scaling from a \pp-like reference value of $\dNdEta \approx 8$ to the measured inclusive \OO value of $\dNdEta \approx 40$~\cite{CMS:2026sai}. The resulting feed-down correction is approximately 2\% at $\pt=0.7\GeV$, increases to about 18\% at $\pt=5\GeV$, and decreases to about 15\% for $\pt>10\GeV$ in both \OO and \pp collisions.

The systematic uncertainty associated with the feed-down correction includes variations of the \PGX and \PGO weighting factors within their experimental uncertainties and contributions from the multiplicity-dependent scaling applied for \OO collisions. The latter is evaluated by considering scenarios with no additional \OO scaling and with a high-multiplicity scaling corresponding to $\dNdEta = 100$ (corresponding to the most head-on collisions). The total uncertainty in feed-down contributions is approximately 2\% for \pp collisions and 4\% for \OO collisions. For \raa, these uncertainties partially cancel between the two collision systems, resulting in a 2\% uncertainty. The systematic uncertainty associated with pileup is estimated by removing the previously described pileup correction.  For \raa, this translates to an uncertainty of about 8\% at low \pt, decreasing to 1--2\% for $\pt>3\GeV$.  Pileup effects partially cancel between \PGL and \PKzS particles in a given collision system, leading to uncertainties smaller than 2\% (1\%) in the \PGL/\PKzS ratio for \OO (\pp) collisions. 

Additional systematic uncertainties arise from imperfect modeling of the detector in simulation. Imperfections in the reconstruction efficiency of daughter tracks are evaluated using a ``tag-and-probe" method~\cite{CMS:2010svw} based on the \PKzS meson.  The observed differences between data and simulation are propagated as a 2\% systematic uncertainty for both \pp and \OO collisions.  These values are added in quadrature for the measurements of \raa and the $\PGL/\PKzS$ ratio. The uncertainty associated with candidate selection is evaluated by varying the selection requirements; this contributes 2--3\% at low \pt, falling below 1\% at higher \pt for the cross section measurements. This uncertainty largely cancels, resulting in an uncertainty of less than 1\% in the \raa measurement.  On the other hand, the candidate selection uncertainties for \PGL and \PKzS particles are added in quadrature for the $\PGL/\PKzS$ ratio because of the difference in decay topologies between the two particles.  The signal extraction uncertainty, estimated by varying the invariant mass fit range, is below 0.5\%. Statistical uncertainties related to finite MC sample size are included as a systematic uncertainty and added in quadrature for the \raa and $\PGL/\PKzS$ ratio measurements. They are below 2\% for most \pt values. The uncertainty in the branching fraction for $\PGL\to\Pp\PGpm$ is 0.8\%~\cite{ParticleDataGroup:2026} and fully cancels for the \raa measurement. Uncertainties related to the \PKzS meson branching fraction, \pt resolution effects, and the modeling of combinatorial background in the invariant mass fit are negligible. Uncertainties resulting from luminosity calibrations are not included in the total systematic uncertainty but are reported separately as a normalization uncertainty fully correlated across the entire \pt range for the spectra and \raa measurements.  Luminosity uncertainties cancel for the $\PGL/\PKzS$ ratio.

Recently, CMS reported the differential cross sections and \raa for inclusive charged hadrons with $\pt>3\GeV$~\cite{CMS:2025bta} using the same datasets as analyzed here.  To facilitate comparison with the strange hadron measurements in this work, which probe lower \pt values, the charged-hadron measurement has been extended to cover the range $0.5<\pt<3\GeV$ using an identical methodology as in Ref.~\cite{CMS:2025bta}.  In this \pt range, the dominant uncertainty in \OO collisions arises from the secondary-track rejection and is approximately 5\%. The leading uncertainty in the \pp measurement comes from differences between data and simulation and is about 2\%.  In this Letter, the results for charged hadrons with $\pt>3\GeV$ are taken from Ref.~\cite{CMS:2025bta}.

\begin{figure}[htp!]
    \centering
    \includegraphics[width=\cmsFigWidth]{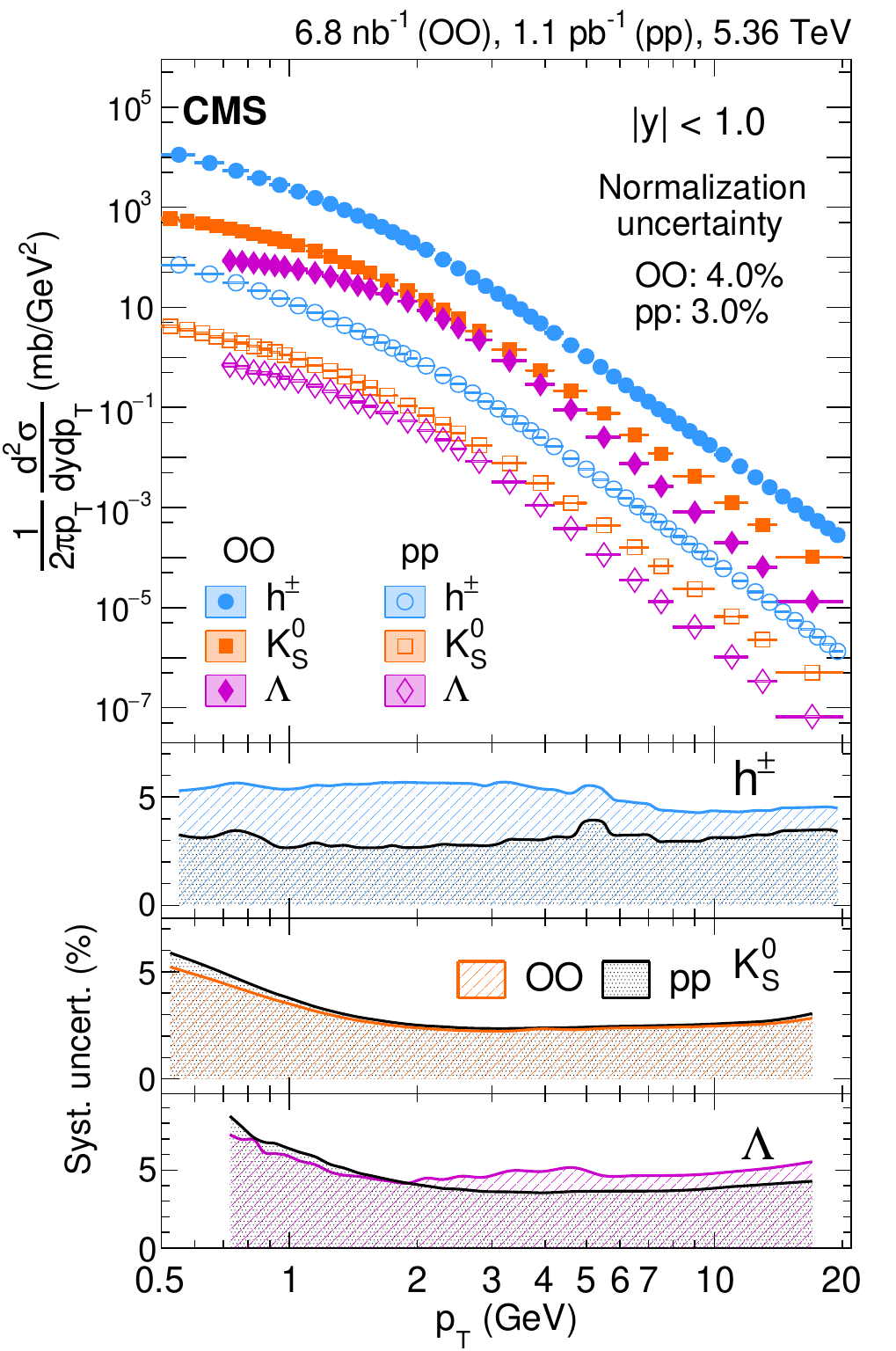}
    \caption{(Upper panel) Invariant differential cross sections of charged hadrons (blue circles), \PKzS mesons (orange squares), and \PGL baryons (magenta diamonds) in \OO (solid markers) and \pp (open markers) collisions as a function of \pt.  The markers show the cross section averaged across the entire bin width.  Statistical uncertainties are shown as error bars and are smaller than the marker size.  (Lower panels) Systematic uncertainties, with normalization uncertainties excluded, for the corresponding cross sections in \OO (colored band) and \pp (black band) collisions.}
    \label{fig:PtSpectra}
\end{figure}

The \pt-differential invariant cross sections of charged hadrons, \PKzS, and \PGL in \pp and \OO collisions are reported in Fig.~\ref{fig:PtSpectra}. A clear dependence on the size of the colliding system is observed, with the \OO values exhibiting larger differential cross sections and a modified spectral shape relative to \pp collisions in the low-\pt region. For \pt above 5\GeV, the spectra exhibit an approximately power-law behavior. Among the different particle species, charged hadrons exhibit the largest cross sections, followed by \PKzS and then \PGL. 

\begin{figure}[ht]
    \centering
    \includegraphics[width=\cmsFigWidth]{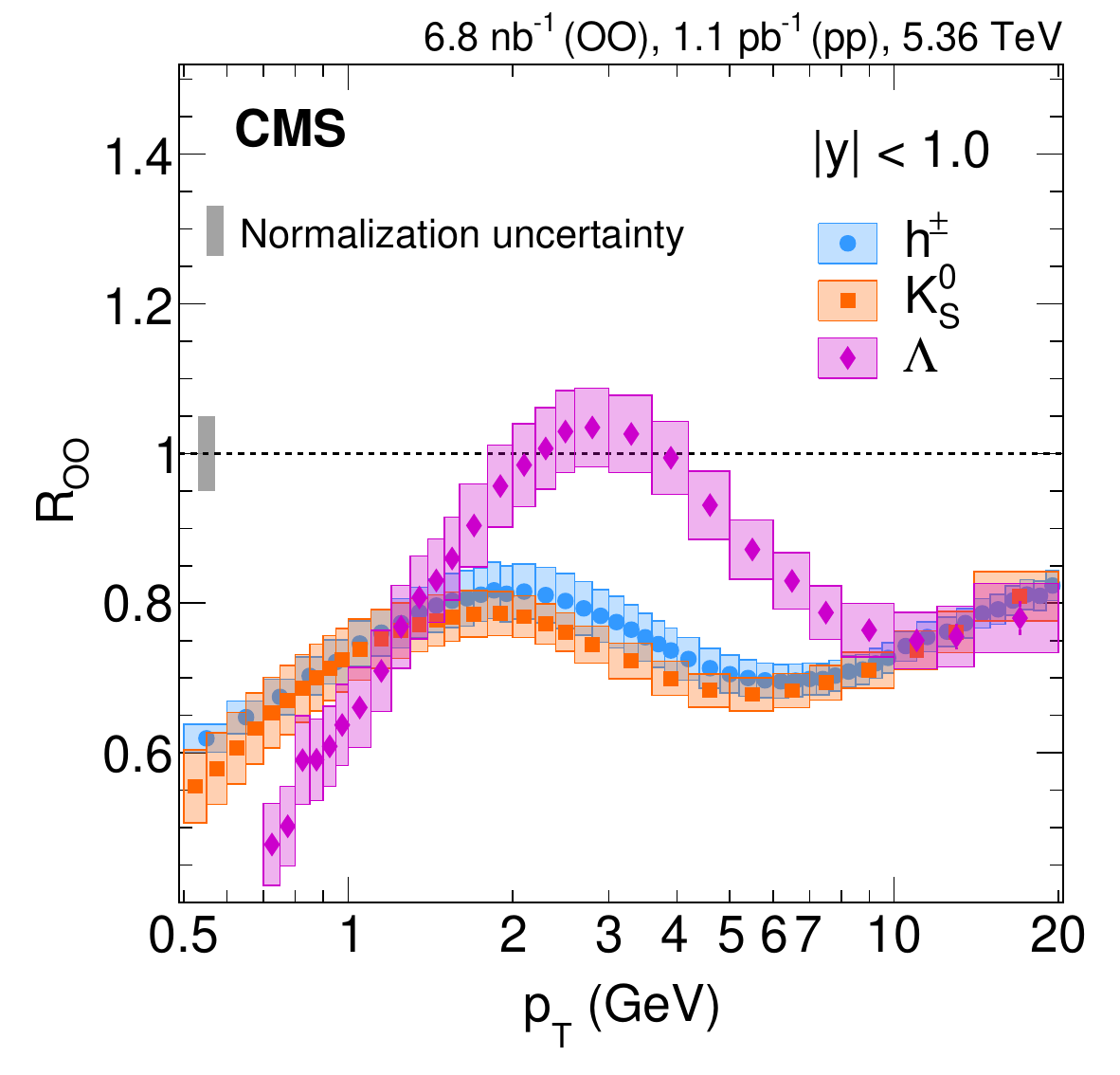}
    \caption{The charged-hadron (blue circles), \PKzS meson (orange squares) and \PGL baryon (magenta diamonds)  nuclear modification factors in \OO collisions.  Statistical uncertainties are shown as error bars, while boxes show systematic uncertainties after excluding the normalization uncertainty. The normalization uncertainty is shown by the gray band around unity.}
    \label{fig:RAA}
\end{figure}

The corresponding \raa values in \OO collisions, denoted as \roo, are shown in Fig.~\ref{fig:RAA}.  For $\pt<8\GeV$, the \PKzS data sit slightly below the \roo for charged hadrons, although the distributions are consistent within the experimental uncertainties.  This is likely because the charged-hadron data includes a small contribution (around 10\%) from protons and antiprotons which may exhibit baryon anomaly effects.  In the intermediate \pt region (2--8\GeV), the \roo of \PGL are enhanced compared to that of \PKzS mesons and charged particles.  This behavior is consistent with a baryon-to-meson enhancement, which is commonly interpreted as a signature of hadronization in a deconfined medium. In the quark coalescence picture~\cite{Fries:2008hs,Minissale:2015zza}, hadrons are formed through the combination of quarks from the medium.  Baryons (containing three constituent quarks) gain larger \pt than mesons (having two constituent quarks), leading to an enhanced baryon yield at intermediate \pt.  In addition, collective radial flow in a hydrodynamically expanding medium~\cite{Heinz:2013th,Jiang:2013gxa} gives heavier hadrons a larger boost in \pt, shifting particles such as the \PGL baryons to higher \pt values than lighter mesons, such as \PKzS.   For $\pt>10\GeV$, the distributions of all three types of particles are nearly identical to each other but are significantly below unity.  As discussed extensively in Ref.~\cite{CMS:2025bta}, this suppression below unity is consistent with nuclear parton distribution function modifications~\cite{Klasen:2023uqj} combined with the effects of highly energetic partons losing energy in a QGP, primarily via gluon radiation~\cite{Huss:2020dwe}.  The convergence of the three particle classes indicates that the difference in strange and light quark masses does not strongly affect these energy loss processes.  All the qualitative trends observed here are consistent with behavior previously seen in lead-lead~\cite{ALICE:2015dtd} and gold-gold collisions~\cite{STAR:2007cqw}, although the magnitudes of the various effects are less pronounced in the smaller \OO system.

\begin{figure}[th]
    \centering
    \includegraphics[width=\cmsFigWidth]{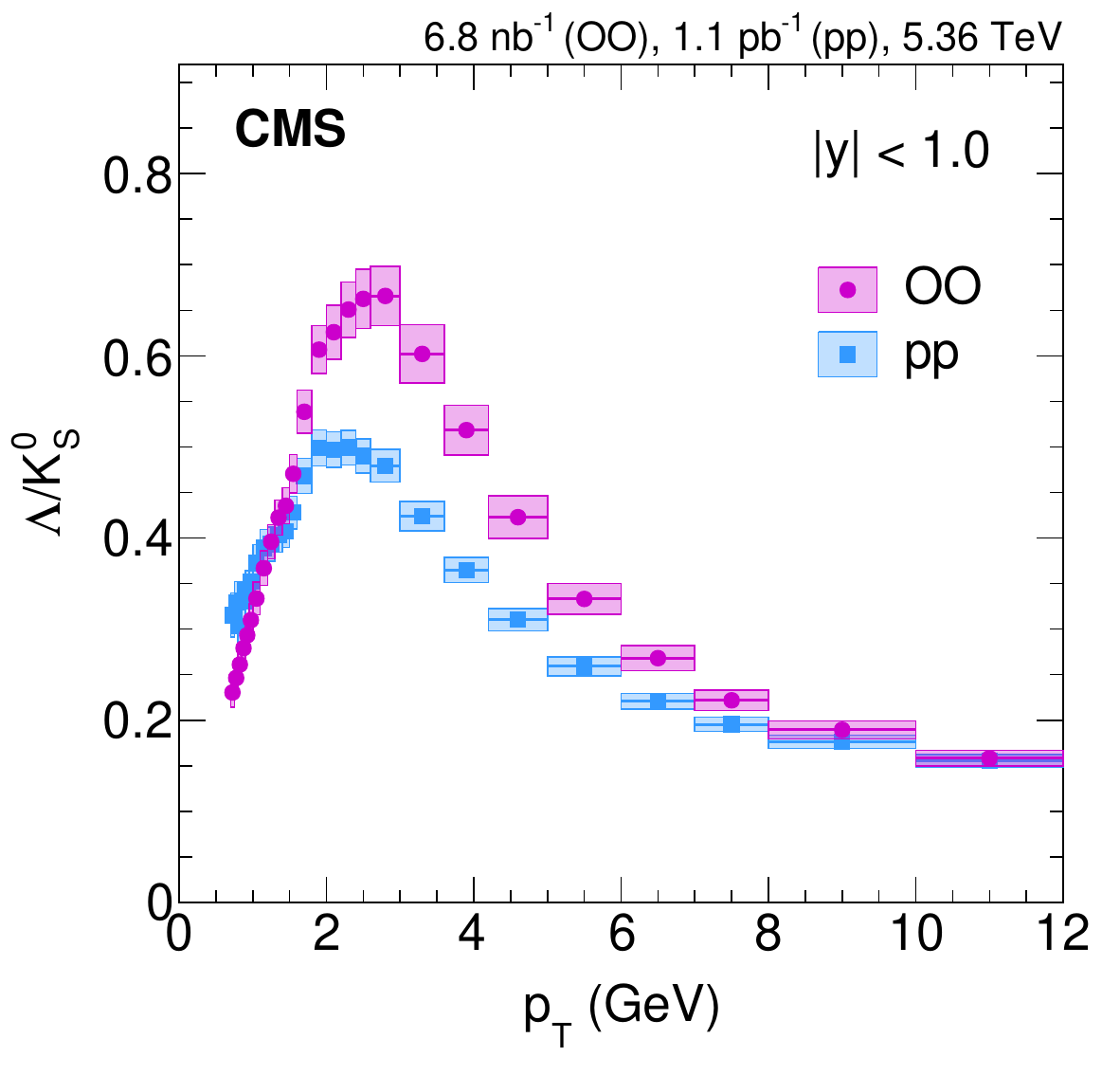}
    \caption{The ratio of the \PGL baryon to \PKzS meson cross sections in \OO (magenta circles) and \pp (blue squares) collisions as a function of \pt. Statistical uncertainties are shown as error bars and are smaller than the marker size. The boxes show the total systematic uncertainties.}
    \label{fig:BtoMRatio}
\end{figure}

Figure~\ref{fig:BtoMRatio} highlights the differences between baryons and mesons by presenting the \PGL/\PKzS cross section ratio in \OO and \pp collisions.  For \pp collisions, the ratio exhibits a rising trend in the range $0.7<\pt<2\GeV$ before reaching a plateau, and a decreasing trend is seen for $\pt>3\GeV$.  Similar behavior has been seen in previous measurements of the same quantity~\cite{ALICE:2025xji}.  In \OO collisions, the same qualitative behavior is observed, but the magnitude of the \PGL enhancement is much larger.  The significance of the difference of the \PGL/\PKzS ratios in \OO and \pp collisions was evaluated by accounting for correlated systematic effects, including those associated with candidate selection, feed-down corrections, and branching fraction uncertainties.  In the intermediate \pt region (2.3--5.2\GeV),  this significance ranges from 5.0 to 6.3 standard deviations, depending on the \pt bin examined.  Thus, this measurement is the first observation of a significant baryon-to-meson enhancement in \OO collisions relative to \pp collisions.  In addition, the peak position of the ratio is shifted from $\pt \approx 2.3\GeV$ in \pp collisions to $\pt \approx 2.8\GeV$ in \OO collisions.  These observations point towards an enhanced baryon-to-meson production consistent with hadronization via quark coalescence and collective flow in a strongly interacting medium. 

In summary, the invariant cross sections of \PKzS and \PGL particles as a function of transverse momentum (\pt) have been measured for the first time in oxygen-oxygen (\OO) collisions at a center-of-mass energy per nucleon pair of 5.36\TeV.  Using a baseline measurement performed with proton-proton (\pp) collisions at the same collision energy, the nuclear modification factor in \OO collisions, \roo, is reported. The \roo of \PKzS mesons is consistent with that of charged hadrons, whereas the \roo of \PGL baryons shows a significant enhancement relative to charged hadrons in the range $2<\pt<8\GeV$.  At higher \pt, both \PKzS and \PGL \roo values are significantly suppressed, likely indicating the presence of parton energy loss effects in a deconfined medium.  An enhancement of the \PGL/\PKzS differential cross section ratio in \OO collisions relative to \pp collisions is observed for $2.3<\pt<5.2\GeV$, with a significance exceeding five standard deviations. This establishes the first observation of a baryon-to-meson enhancement in \OO collisions and indicates modified hadron production suggesting quark coalescence and collective radial flow in a strongly interacting deconfined medium.  

\begin{acknowledgments}
We congratulate our colleagues in the CERN accelerator departments for the excellent performance of the LHC and thank the technical and administrative staffs at CERN and at other CMS institutes for their contributions to the success of the CMS effort. In addition, we gratefully acknowledge the computing centers and personnel of the Worldwide LHC Computing Grid and other centers for delivering so effectively the computing infrastructure essential to our analyses. Finally, we acknowledge the enduring support for the construction and operation of the LHC, the CMS detector, and the supporting computing infrastructure provided by the following funding agencies: SC (Armenia), BMFWF and FWF (Austria); FNRS and FWO (Belgium); CNPq, CAPES, FAPERJ, FAPERGS, and FAPESP (Brazil); MES and BNSF (Bulgaria); CERN; CAS, MoST, and NSFC (China); MINCIENCIAS (Colombia); MSES and CSF (Croatia); RIF (Cyprus); SENESCYT (Ecuador); ERC PRG and PSG, TARISTU24-TK10 and MoER TK202 (Estonia); Academy of Finland, MEC, and HIP (Finland); CEA and CNRS/IN2P3 (France); SRNSF (Georgia); BMFTR, DFG, and HGF (Germany); GSRI (Greece); MATE and NKFIH (Hungary); DAE and DST (India); IPM (Iran); SFI (Ireland); INFN (Italy); MSIT and NRF (Republic of Korea); MES (Latvia); LMTLT (Lithuania); MOE and UM (Malaysia); BUAP, CINVESTAV, CONACYT, LNS, SEP, and UASLP-FAI (Mexico); MOS (Montenegro); MBIE (New Zealand); PAEC (Pakistan); MSHE, NSC, and NAWA (Poland); FCT (Portugal); MESTD (Serbia); MICIU/AEI and PCTI (Spain); MOSTR (Sri Lanka); Swiss Funding Agencies (Switzerland); MST (Taipei); MHESI (Thailand); TUBITAK and TENMAK (T\"{u}rkiye); NASU (Ukraine); STFC (United Kingdom); DOE and NSF (USA).
\end{acknowledgments}\section*{Data availability} Release and preservation of data used by the CMS Collaboration as the basis for publications is guided by the  \href{https://doi.org/10.7483/OPENDATA.CMS.1BNU.8V1W}{CMS data preservation, re-use and open access policy}.
\bibliography{auto_generated}

\cleardoublepage \appendix\section{The CMS Collaboration \label{app:collab}}\begin{sloppypar}\hyphenpenalty=5000\widowpenalty=500\clubpenalty=5000\cmsinstitute{Yerevan Physics Institute, Yerevan, Armenia}
{\tolerance=6000
A.~Belyaev\cmsorcid{0000-0003-1692-1173}, A.~Gevorgyan\cmsorcid{0000-0003-2751-9489}, A.~Hayrapetyan, A.~Tumasyan\cmsAuthorMark{1}\cmsorcid{0009-0000-0684-6742}
\par}
\cmsinstitute{Institut f\"{u}r Hochenergiephysik, Vienna, Austria}
{\tolerance=6000
P.S.~Hussain\cmsorcid{0000-0002-4825-5278}, M.~Sonawane\cmsorcid{0000-0003-0510-7010}
\par}
\cmsinstitute{Marietta Blau Institute for Particle Physics, Vienna, Austria}
{\tolerance=6000
W.~Adam\cmsorcid{0000-0001-9099-4341}, L.~Benato\cmsorcid{0000-0001-5135-7489}, T.~Bergauer\cmsorcid{0000-0002-5786-0293}, M.~Dragicevic\cmsorcid{0000-0003-1967-6783}, S.~Gundacker\cmsorcid{0000-0003-2087-3266}, A.K.~Guven\cmsorcid{0009-0004-5670-5138}, M.~Jeitler\cmsAuthorMark{2}\cmsorcid{0000-0002-5141-9560}, N.~Krammer\cmsorcid{0000-0002-0548-0985}, A.~Li\cmsorcid{0000-0002-4547-116X}, D.~Liko\cmsorcid{0000-0002-3380-473X}, M.~Matthewman, A.~Pfeiffer\cmsorcid{0000-0001-5328-448X}, J.~Schieck\cmsAuthorMark{2}\cmsorcid{0000-0002-1058-8093}, R.~Sch\"{o}fbeck\cmsAuthorMark{2}\cmsorcid{0000-0002-2332-8784}, M.~Shooshtari\cmsorcid{0009-0004-8882-4887}, N.~Van~Den~Bossche\cmsorcid{0000-0003-2973-4991}, W.~Waltenberger\cmsorcid{0000-0002-6215-7228}, C.E.~Wulz\cmsAuthorMark{2}\cmsorcid{0000-0001-9226-5812}
% ...
\par}
$^{1}$Also at Yerevan State University, Yerevan, Armenia\\
$^{2}$Also at Technische Universit\"{a}t Wien, Vienna, Austria\\
% ...
\end{sloppypar}
\end{document}